%% file: gnr_hbn_draft_06.tex
\RequirePackage{snapshot}
\documentclass[prb,aps,showpacs,twocolumn,unsortedaddress]{revtex4}

\input{mycommands}

\begin{document}

\title{Electronic structure of graphene-nanoribbons on hexagonal boron nitride.}

\author{Yohanes S. Gani$^1$, D.S.L. Abergel$^2$\footnote{Present address: Nature Physics, 4 Crinan Street, London, N1 9XW, UK}, Enrico Rossi$^1$}
\affiliation{
             $^1$Department of Physics, William \& Mary, Williamsburg, VA 23187, USA\\
             $^2$Nordita, KTH Royal Institute of Technology and Stockholm University, Roslagstullsbacken 23, 10691 Stockholm, Sweden
            }
\date{\today}

   
\begin{abstract}
Hexagonal boron nitride is an ideal dielectric to form two-dimensional heterostructures due to the fact that 
it can be exfoliated to be just few atoms thick and its a very low density of defects. 
By placing graphene nanoribbons on high quality hexagonal boron nitride it is possible to create 
ideal quasi one dimensional (1D) systems with very high mobility. The availability of high quality one-dimensional
electronic systems is of great interest also given that when in proximity to a superconductor they can 
be effectively engineered to realize Majorana bound states. In this work we study how a boron nitride substrate 
affects the electronic properties of graphene nanoribbons. We consider both armchair and zigzag nanoribbons.
Our results show that for some stacking configurations the boron nitride can significantly 
affect the electronic structure of the ribbons. 
In particular, for zigzag nanoribbons, due to the lock between spin and sublattice degree of freedom at the edges,
the hexagonal boron nitride can induce a very strong spin-splitting of the spin polarized, edge sates. 
We find that such spin-splitting can be as high as 40~meV.
\end{abstract}


\maketitle


\section{Introduction}

Graphene nanoribbons (GNRs)
\cite{Nakada1996,Ezawa2006,Barone2006,Fujita1996,kn:son2006,kn:yang2007,Dutta2010}
are almost ideal 1D electronic systems: they are only one atom thick and their width can be just few atoms.
Recent advances in bottom-up synthesis using molecular precursors
allow to control with atomic precision the width and the edges' morphology og GNRs.
\cite{Cai2010,Pascal2016,Narita2013}.
These developments make GNRs very promising as basal elements for the realization
of quasi-1D systems and 1D topological states~\cite{Oliver2018,Rizzo2018}. The particular advantage of GNRs toward this goal are:
(i)   almost ideal 1D character,
(ii) scalable synthesis and layout to create networks of quasi 1D channels,
(iii)  tunability of their electronic properties via edge and width engineering.
Interest in 1D electronic systems has recently increased substantially
given that to date the most successful and promising approaches to realize non-abelian
electronic states, such as Majoranas,
rely on the availability of 1D devices~\cite{Lutchyn2018} 
of high quality (ideally disorder free)~\cite{Potter2011, Potter2011a, Lobos12, Lutchyn2012, Sau2013, Hui2015, Cole2016, Liu2017}.
The ultimate 1D nature of GNRs
and therefore large energy separation between their 1D subbands makes them in many respects
ideal for the realization of 1D devices.

To be able to use GNRs to realize states like Majoranas the GNRs have to be of very high quality,
i.e to have a very low level of disorder.
In recent years high quality hexagonal boron nitride, hBN, has emerged as the ideal dielectric to realize graphene-based
heterostructures~\cite{dean2010,yang2013,xue2011,ponomarenko2011,yankowitz2012}.
This is due to the fact that hBN has a large band gap, a very low density of impurities and crystal defects,
and it can be exfoliated to be only few atoms thick.
Because of the extreme low impurity density of hBN,
graphene devices in which hBN is the dielectric substrate have electron's mobilities
orders of magnitude larger than graphene devices on other substrate, such as, for example, 
silicon dioxide~\cite{Novoselov2004,Nomura2006,Tan2007,Hwang2007,dassarma2010,DasSarma2011}.
One additional important consequence of having a substrate with low disorder, is that
in systems like graphene and bilayer graphene, it also reduces the carrier density inhomogeneities that, especially close
to the Dirac point or in the presence of a small band gap, can be very large and significantly
modify the electronic properties of the graphene-based device~\cite{Adam2007,fogler2007,Rossi2008,Polini2008,adam2008,fogler2009,rossi2009,Rossi2011,Li2011,Abergel2012,abergel2013,Li2012,jzhang2013,jzhang2014,rodriguez2014,rodriguez2017}.
Imaging experiments have directly shown that the use of hBN as a substrate instead of silicon oxide
greatly reduces the amplitude of the disorder-induced carrier density 
inhomogeneities~\cite{xue2011,yankowitz2012,Lu2016}.

For all the reasons stated above it is natural to use hBN as a substrate for graphene nanoribbons.
However, it has been shown both theoretically~\cite{song2013,jung2014,song2015,jung2015,jung2017}
and experimentally~\cite{yankowitz2012,britnell2012,hunt2013,ponomarenko2013,dean2013}
that hBN can qualitatively affect the band structure of graphene.
This is due to the fact that in graphene-hBN devices, because hBN has a lattice constant that is only 1.8\% larger than graphene's,
there can be region tens of nanometer wide in which the graphene layer is in register with the hBN lattice~\cite{woods2014}
and therefore have its sublattice symmetry broken given that in hBN the A and B sublattices create different electrostatic potentials.
Given that GNRs are typically only few atoms wide we should expect that hBN can qualitatively modify their band structure.
In order to be able to use hBN to increase the quality of GNRs to realize almost ideal 1D electronic systems,
it is therefore necessary to understand how hBN can affect the spectrum of GNRs.
In this work we study how hBN modifies the band structure of GNRs. We study different types of GNRs and consider
different (commensurate) stackings between the GNRs and hBN. We find that hBN can cause qualitative changes to
the band structure of GNRs and that these changes can be tuned by selecting the stacking configuration.
The effects are most dramatic for zig-zag graphene nanoribbons (GNRs): for such ribbons  hBN in general induces
a spin splitting of the conduction band (CB) and valence band (VB). We also find that the sign of such spin-splitting
can be changed simply by changing, via a rigid shift, the stacking between the ZGNR and hBN.

The paper is organized as follows: 
in Sec.~\ref{sec:method} we present the theoretical method that we use and a brief review of the electronic structure for isolated GNRs and hBN,
in Sec.~\ref{sec:results} we present the results for the band structure of GNR-hBN heterostructures, 
and in Sec.~\ref{sec:conclusions} we provide our conclusions.

\section{Method}
\label{sec:method}

\begin{figure}[htb]
 \begin{center}
  \includegraphics[width=0.85\columnwidth]{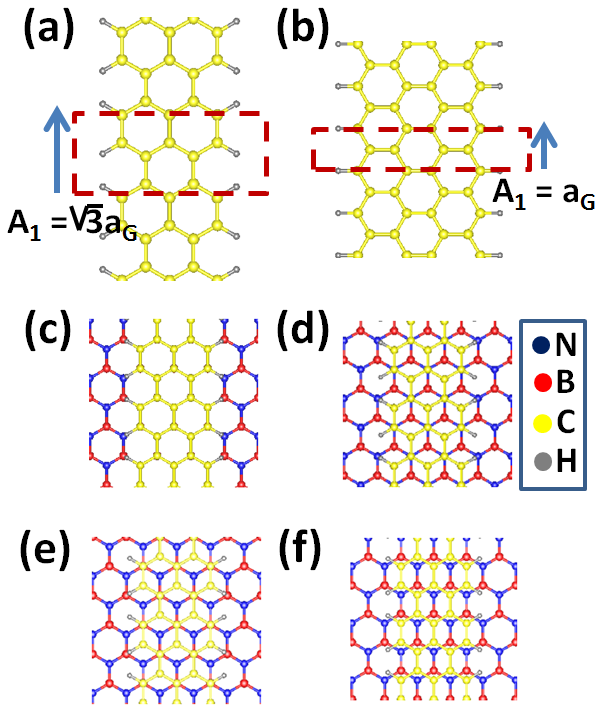}
  \caption{
           Atoms layout for AGNR (a), and ZGNR (b).
           The dashed lines identify the primitive cells. 
           (c), (d), (e), (f) possible stacking configurations between a GNR and hBN: 
           AA,  $\mathrm{AB_N}$, $\mathrm{AB_B}$, and $\mathrm{A_{br}}$, respectively.
         } 
  \label{fig:gnrs}
 \end{center}
\end{figure}

Graphene is a one-atom thick layer of carbon atoms arranged in an hexagonal structure~\cite{Novoselov2004,Zhang2005,CastroNeto2009,DasSarma2011}.
In graphene the carbon-carbon distance, $a$, is 1.46~\AA.
The hexagonal structure is best described as a triangular lattice with lattice constant $a_G=\sqrt{3}a$ and a basis with two sites, A and B.
The atoms at sites A form the A-sublattice and the atoms at the B sites form the B-sublattice. 
In graphene the A and B sites are both occupied by carbon atoms and so we have sublattice symmetry.
Graphene nanoribbons can be obtained by etching graphene along particular directions~\cite{Campos2009}.
More recently, GNRs have been produced via bottom-up synthesis~\cite{Cai2010,Pascal2016,Narita2013}, a fabrication technique
that allows to control with atomic precision the width of the ribbon and the shape of their edge and therefore their electronic properties.
Depending on their edges we can identify two types of GNRs: armchair GNRs (AGNRs), Fig.~\ref{fig:gnrs}~(a), 
in  which the edges look like a sequence of armchairs, and zigzag GNRs (ZGNRs), Fig.~\ref{fig:gnrs}~(b), 
in which the edges have a zigzag pattern.
It is customary to refer to the width of an AGNR via the number $N$ across the transverse direction
of carbon-carbon dimers aligned along the longitudinal direction. For ZGNRs the width is denoted by the number $N$
of zigzag chains. For the remainder it is important to notice that the unit cell of AGNRs and ZGNRs is different,
as shown in  Fig.~\ref{fig:gnrs}. Let $\aagnr$ be the nanoribbon lattice constant. For AGNRs $\aagnr = \sqrt{3}a_G$,
for ZGNRs  $\azgnr=a_G$.

The heterostructures that we study are formed by a graphene nanoribbon (armchair of zigzag) placed on hBN.
Figure~\ref{fig:gnrs}~(c-f) show some examples GNR-hBN structures.
In hBN the sublattice A (B) is occupied by boron (nitrogen) atoms, or vice versa.
The fact that the A and B sites are not equivalent in hBN in Fig.~\ref{fig:gnrs}, and all the figures in the remainder
of this work, is denoted by the fact that they are shown in different colors.
In all the results presented in the remainder, to avoid the effects due to dangling bonds, we assume the edges of the GNRs to be terminated
by hydrogen atoms, showing in light grey in Fig.~\ref{fig:gnrs}. 
It is helpful to name the particular stackings shown in Fig.~\ref{fig:gnrs}.
Figure~\ref{fig:gnrs}~(c) shows the case in which the ribbon and the hBN are in the AA stacking configuration, i.e.
the case in which the GNR's sublattice A (B) is directly above the sublattice A (B) of hBN.
In the $AB_N$ ($AB_B$) stacking the sublattice A (B) of the GNR is in register with the sublattice occupied by the nitrogen (boron) atoms of 
the substrate,~Fig.~\ref{fig:gnrs}~(d) (Fig.~\ref{fig:gnrs}~(e)).
In the bridge-stacking configuration, \abr, 
the carbon-carbon links of the GNR cross the boron-nitrogen links of the substrate,~Fig.~\ref{fig:gnrs}~(f).

\begin{figure}[htb]
 \begin{center}
  \includegraphics[width=0.9\columnwidth]{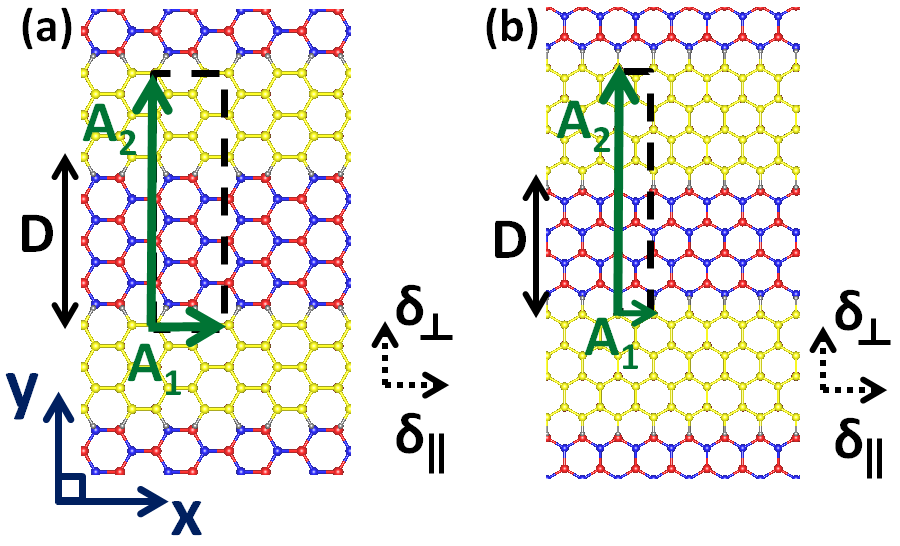}
  \caption{ 
           Sketch of the supercell used for the DFT calculation. $D$ is the distance between
           ribbons, and $\delta_\perp$ ($\delta_\parallel$) denote transverse (longitudinal) shifts 
           of the GNR with respect to the hBN substrate away from AA stacking.
         } 
  \label{fig:scheme}
 \end{center}
\end{figure} 

The tight-binding model is a computationally very efficient method that has been used to obtain the 
band structure of GNRs~\cite{Nakada1996,White2008,Zheng2007,Fertig2006} and related systems. 
However, to get accurate results, even qualitatively, using the tight-binding model requires a fine tuning of its parameters 
that can only be achieved by comparing the tight-binding model's results to the ones obtained using 
density functional theory (DFT) approaches~\cite{kn:son2006,Raza2008,Sahu2008,Barone2006,Lee2009,kn:yang2007}.
For the case of isolated AGNR this is exemplified by the 
fact that the simple nearest neighbor tight binding model with constant hopping parameter
for the case in which $N=3n-1$ return a gapless band-structure whereas DFT shows the presence of finite band-gap~\cite{kn:son2006,kn:yang2007}.
The main reason for such discrepancy is that, due to the finite width of the ribbon, the hopping parameter entering
the tight-binding model should not be taken to be constant across the ribbon's width~\cite{kn:son2006}
and hopping processes beyond next-neighbor should also be included~\cite{White2008}. 
For ZGNRs the simple tight-binding model predicts a gapless band structure, due to the presence of edge modes, a fact
that is not affected by the variation of the hopping parameter across the ribbon. However, also for ZGNRs 
the result of the simple tight-binding model are qualitatively incorrect if one does not include the effect of
the exchange part of the Coulomb interaction.
The exchange interaction causes ZGNRs to have an insulating ground state with ferromagnetic order along the edges and
antiferromagnetic order between the two edges, effect that is correctly captured by ab-initio calculations~\cite{Lee2005,son2006,Kan2012}.

For these reasons, in this work  
we obtain the electronic structure of all the systems via ab-initio density functional theory calculations
using the Quantum Espresso package~\cite{Giannozzi2009}. We use ultrasoft potentials and a plane-waves basis with periodic boundary conditions.

We denote as $x$ the axis along the longitudinal direction of the GNR, as $z$ the axis perpendicular to the heterostructure plane
and as $y$ the axis in the GNR plane perpendicular to both $x$ and $z$, as shown in Fig.~\ref{fig:scheme}. $\delta_\parallel$ ($\delta_\perp$) denotes a shift along the $x$ ($y$) direction between the GNR and the substrate.
In order to simulate a heterostructure with an isolated GNR we need to
use a supercell large enough to minimize artificial interference effects arising from the 
periodic boundary conditions. 
We find that for supercell sizes $D>9 a_G$ finite size effects are negligible and  
do not affect the electronic structure of the GNR.
In the direction perpendicular to the plane of the GNR-hBN heterostructure we insert a ``vacuum layer'' 10~$\mathrm{\AA}$ thick.

The electron exhange and correlation are calculated by implementing the generalized gradient approximation (GGA) functional of Perdew-Burke-Ernzerhof (PBE) \cite{Perdew1996}.
For AGNR hybrid systems the Brillouin zone (BZ) integration is performed by generating a uniform 12x12x1
mesh of k points using the Monkhorst-Pack procedure.
For ZGNR hybrid systems we use the same procedure using 16x16x1 mesh. The cut off energy wavefunction and charge densities are set to be 50 Ry and 400 Ry, respectively, ensuring the convergence of the total energy. 
To be able to compare the effect of different stacking configurations we keep the
interlayer distance $d$ fixed. We conservatively set $d=3.5 \mathrm{\AA}$ considering that the modifications of the 
GNR electronic structure due to the presence of the substrate are stronger for smaller values of $d$.
Changes in $d$ do not change qualitatively the results that we present in remainder. 

We limit ourselves to the case when the stacking between the nanoribbon and hBN is commensurate . 
We assume that the 1.8\% lattice mismatch between the graphene
nanoribbon and hBN can be neglected given the small size of the system and the fact that
in graphene-hBN heterostructures it has been shown that graphene and hBN lattices can be in commensurate
stacking configurations over regions tens of nanometers wide. 
wide~\cite{woods2014}.

\section{Results}
\label{sec:results}

\begin{figure}[htb]
 \begin{center}
  \includegraphics[width=\columnwidth]{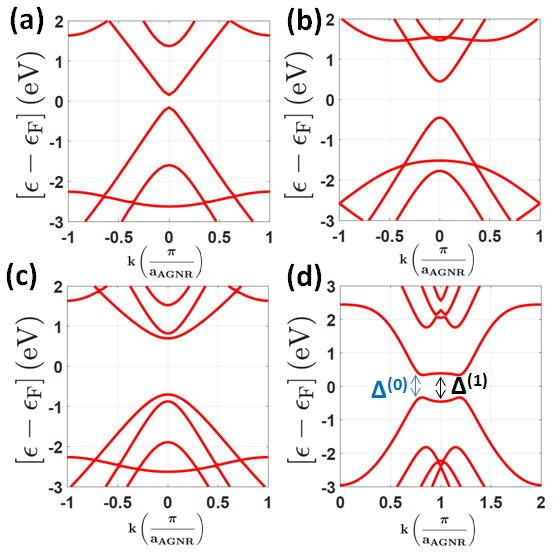}
  \caption{            
           (a) Band structure for an AGNR with N=3n-1=5, (n=2). 
           (b) Band structure for an AGNR with N=3n=6.
           (c) Band structure for an AGNR with N=3n+1=7.    
           (d) Band structure for a a  ZGNR with N=4. 
         } 
  \label{fig:isolated-GNRs}
 \end{center}
\end{figure}

\begin{figure}[htb]
 \begin{center}
  \includegraphics[width=4.5cm]{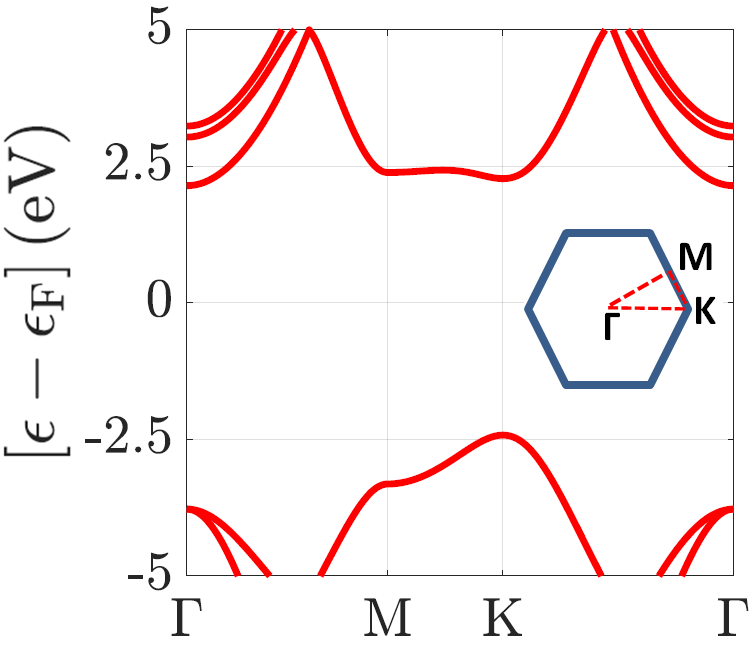}
  \caption{            
           Low energy band structure of hBN. The inset shows the Brillouin Zone.
         } 
  \label{fig:hBN}
 \end{center}
\end{figure}

In this section we present our results.
To better understand the results for the GNR-hBN heterostructures it is helpful to briefly review the electronic structure of isolated GNRs and hBN.
Figure~\ref{fig:isolated-GNRs} shows the low-energy band structure of isolated GNRs obtained using DFT, see Sec.~\ref{sec:method}. Figures~\ref{fig:isolated-GNRs}~(a)-(c) show the band structure for AGNRs with width $N=3n-1$,  $N=3n$, $N=3n+1$, respectively for the case 
when $n=2$. As discussed in Sec.~\ref{sec:method} for all three cases we have a gapped band structure.
Figure~\ref{fig:isolated-GNRs}~(d) shows the band structure for a ZGNR of width $N=4$. Notice that for a ZGNR the low energy states
are located at the edge of the 1D BZ ($k=\pi/a_{\rm ZGNR}$), and the gap due to the antiferromagnetic ordering,
decreases with the width of the ribbon.
Here, and in the remainder, $\Deltaz$ denotes the direct band gap and $\Deltao$ the energy splitting for $k=\pi/\azgnr$. 
Figure~\ref{fig:hBN} shows the low energy band structure of hBN. 

%

\subsection{AGNR-hBN heterostructures}

\begin{figure}[htb]
 \begin{center}
  \includegraphics[width=\columnwidth]{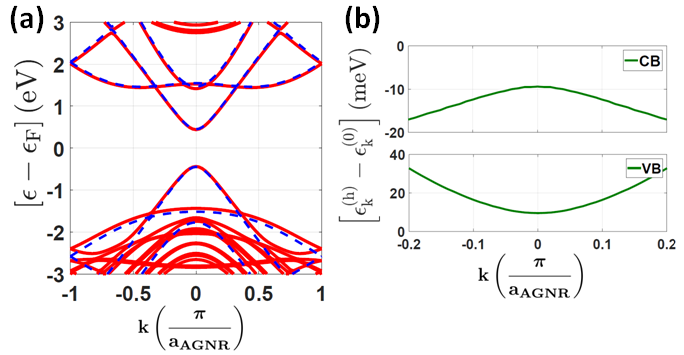}
  \caption{            
           (a) Bands of a AGNR-hBN heterostructure for a ribbon with $N=6$ placed on hBN in the AA stacking configuration.
           The dashed lines show the spectrum of the isolated GNR and the solid lines
           the spectrum of the heterostructure.
           (c), (d) energy shift as a function of $k$ of the CB, and VB, respectively. 
         } 
  \label{fig:agnr-hbn-AA}
 \end{center}
\end{figure}

In this section we present the results for heterostructures formed by AGNR and hBN. Figure~\ref{fig:agnr-hbn-AA}~(a) shows the low-energy band-structure of a AGNR-hBN heterostructure in the AA stacking configuration: here and in the remainder the dashed lines show the spectrum of the isolated GNR and the solid lines
the spectrum of the heterostructure. We see that for this configuration the presence of the hBN
does not modify significantly the spectrum of the GNR. 
Figure~\ref{fig:agnr-hbn-AA}~(b) shows the shift in energy of the ribbon valence and conductance band due to the presence
of the hBN: we see that for this configuration the variation in energy is of the order of 15~meV
close to the $k=0$ point and slowly increases (in absolute value) as we move away from $k=0$.

\begin{figure}[htb]
 \begin{center}
 \includegraphics[width=\columnwidth]{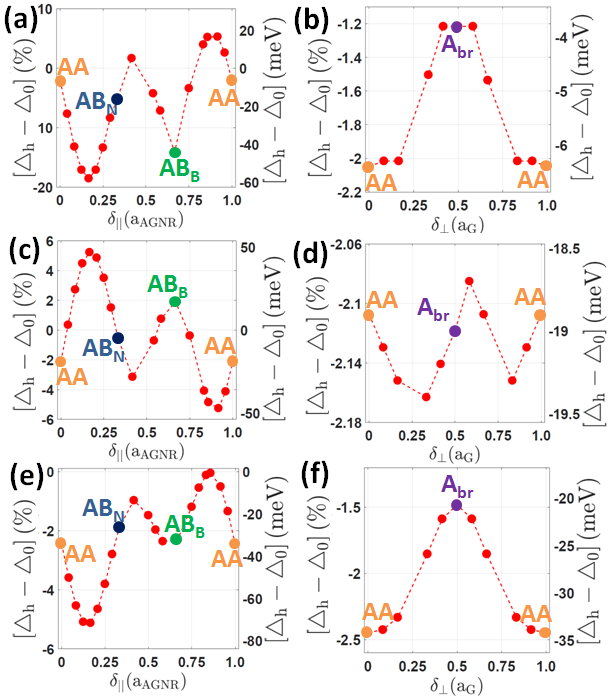}
  \caption{           
           Evolution of the band gap of a AGNR (with $n=2$) placed on hBN as a function of shift away from AA stacking.
           The left panels show the results for a shift along $\delta_{\parallel}$, the right panels for shifts along $\delta_{\perp}$.
           The different rows show the results for different widths of the ribbon:
           the first row (panels (a) and (b)) show the results for the case when  $N=3n-1=5$, 
           the second (panels (c) and (d))  for the case when  $N=3n=6$,
           and the last (panels (e) and (f))  for the case when  $N=3n+1=7$.
         } 
  \label{fig:AGNR-hBN-shifts}
 \end{center}
\end{figure} 

To study how differences in stacking affect the spectrum we studied the effect of a shift away from the AA configuration
in the longitudinal and transverse direction. The relative change of the ribbon's band gap $\Delta_r\df(\Delta_h-\Delta_0)/\Delta_0$
where $\Delta_h$ is the band gap of the GNR-hBN heterostructure, 
can be used to show in a compact way the effect.
The results are shown in Fig~\ref{fig:AGNR-hBN-shifts} for the three classes
of AGNRs: $N=3n-1$, $N=3n$, $N=3n+1$ where, as in the remainder of this work, we have taken $n=2$.
We see that a shift in the perpendicular direction has only a minor effect: the relative change is at most of the order of 2\%.
We also observe that
the highest increase of the band gap due to $\delta_\perp$ is obtained when the shift
results in the \abr configuration for $N=3n-1$ and $N=3n+1$ AGNRs and very close to it for $N=3n$ AGNRs.

\begin{figure}[htb]
 \begin{center} 
  \includegraphics[width=\columnwidth]{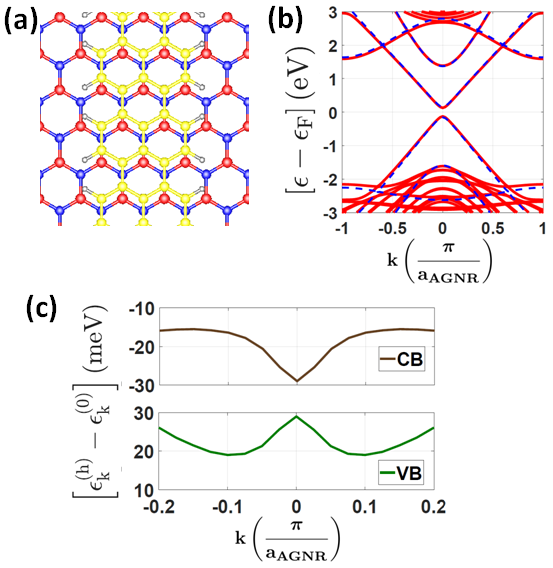}
  \caption{           
           (a) Stacking configuration for a AGNR-hBN system corresponding to the maximum gap change shown
           in Fig.~\ref{fig:AGNR-hBN-shifts}~(a) ($N=5$) corresponding to $ \delta_{||} = 0.16 \left(\frac{1}{a}\right)$.
           (b) Bands for the stacking configuration shown in (a) (the dashed lines show the bands for the isolated ribbon).
           (c) The top panel shows the difference at small $k$'s between the heterostructure's conduction band, CB,
           and the isolated ribbon's CB for the stacking configuration shown in (a). 
           The bottom panel show the difference between the VBs.
         } 
  \label{fig:AGNR-hBN-max}
 \end{center}
\end{figure}

The shift in the longitudinal direction has a stronger effect than $\delta_\perp$. 
By varying $\delta_\parallel$ we can obtain the $AB_N$ and $AB_B$ configurations. 
Figure~\ref{fig:AGNR-hBN-shifts} shows that for all the three types of ribbons $\Delta_r(\delta_\parallel)$ has an extremum 
when the $AB_B$ configuration is realized.
For most cases a shift in the longitudinal direction can induce a change of the
band gap of the order of 6\% or less, however, for the case when $N=3n-1$, i.e. for the class of AGNRs for 
which $\Delta_0$ is the smallest (zero using a tight binding model with uniform hopping parameters)
a shift in the longitudinal direction away from the AA stacking can lead to a configuration for which 
the band gap is reduced by 20\%, i.e. about 60~meV in absolute terms.
Figure~\ref{fig:AGNR-hBN-max} shows the atoms arrangement for this configuration, and the corresponding low-energy band-structure.
We see that for this stacking the nitrogen
atoms are located midway under the longitudinal C-C bonds.
%

\subsection{ZGNR-hBN heterostructures}

%
\begin{figure}[htb]
 \begin{center}
  \includegraphics[width=\columnwidth]{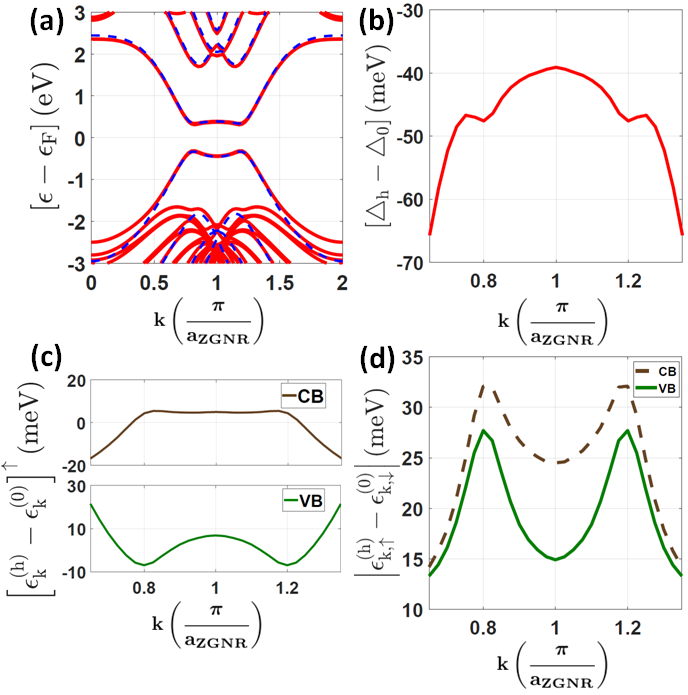}
  \caption{          
           Results for a ZGNR with $(N=4)$ placed on hBN in the AA stacking configuration. 
           (a) Band structure, 
               the dashed lines show the bands for the isolated ZGNR.    
           (b) Difference, for $k$ close to $\pi/a_\mathrm{ZGNR}$, between the band gap of the hBN-ZGNR heterostructure, $\Delta_h$,
               and the band gap of the isolated ZGNR $\Delta_0$.
           (c) The top panel shows the difference for $k$ close to $\pi/a_\mathrm{ZGNR}$ between the ZGNR-hBN heterostructure's CB
               and the isolated ribbon's CB for the AA stacking configuration. 
               The bottom panel show the difference between the VBs.
           (d) Spin splitting as a function of $k$ for the ZGNR-hBN heterostructure's CB and VB.
         } 
  \label{fig:ZGNR-hBN-AA}
 \end{center}
\end{figure} 

\begin{figure}[htb]
 \begin{center}
  \includegraphics[width=\columnwidth]{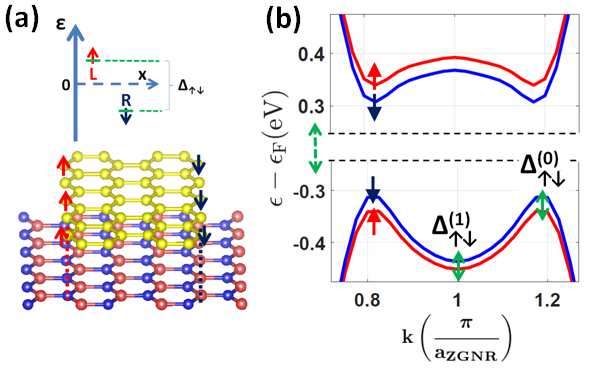}
  \caption{
    (a) Sketch of a ZGNR placed on hBN in the AA stacking configuration. The arrows at the edges of the ZGNR show
        the spin polarization of the edge modes. 
    (b) Enlargement of the VB and CB to show the spin splitting due to the presence of hBN.
           } 
  \label{fig:sketch01}
 \end{center}
\end{figure}

We now consider ZGNR-hBN heterostructures. 
Figure~\ref{fig:ZGNR-hBN-AA} shows the low energy spectrum of a ZGNR-hBN heterostructure for the case of AA stacking.
Analogously to what we find for AGNR-hBN we see that for this configuration the effect of the hBN on the band gap
is small: the conduction and valence bands around $k=\frac{\pi}{a_{\rm ZGNR}}$ are shifted by 10-20~meV, Fig.~\ref{fig:ZGNR-hBN-AA}~(c).
However, the presence of hBN causes an important qualitative modification of the band structure:
it induces a spin splitting of the valence and conduction bands, see Fig.~\ref{fig:ZGNR-hBN-AA}~(d).
This is due to the locking between spin and sublattice degrees of freedom for the edge states and
the fact that the presence of hBN breaks the GNR sublattice symmetry.
For ZGNRs the left (right) edge state has spin polarization up (down) while at the same
time the atoms forming the left (right) edge belong to the A (B) sublattice (or viceversa).
As Fig.~\ref{fig:sketch01} shows the presence of hBN breaks the sublattice symmetry
and therefore the degeneracy of the states due to this symmetry. In a ZGNR, 
the breaking of the sublattice symmetry therefore causes a spin splitting of the edge states,
for which spin and sublattice degrees of freedom are locked.
The effect of the presence of hBN on the band structure of ZGNR is similar to the
effect of an electric field applied along the transverse direction of a ZGNR. It was shown that for large enough 
transverse electric fields a ZGNR can be driven into an ideal half-metallic state~\cite{Son2006b,yu2013}.
For the case of a ZGNR placed on hBN the difference in electrostatic potential between the ZGNR's atoms on the two
different edges is not due to an external electric field but the fact that they are located
above different atoms of the layer forming the substrate.
The results of Figure~\ref{fig:ZGNR-hBN-AA}~(d) show that hBN, and any substrate that break the sublattice symmetry of graphene,
can be used to spin split the edge modes of a ZGNR.
We can conclude that in ZGNR-hBN heterostructures we can break 
the spin-degeneracy without having to introduce an external magnetic field and explicitly
breaking the time reversal symmetry.
It is interesting to see if such an effect can be maximized by tuning the stacking configuration and the width of the ZGNR.

\begin{figure}[h]
 \begin{center}
  \includegraphics[width=\columnwidth]{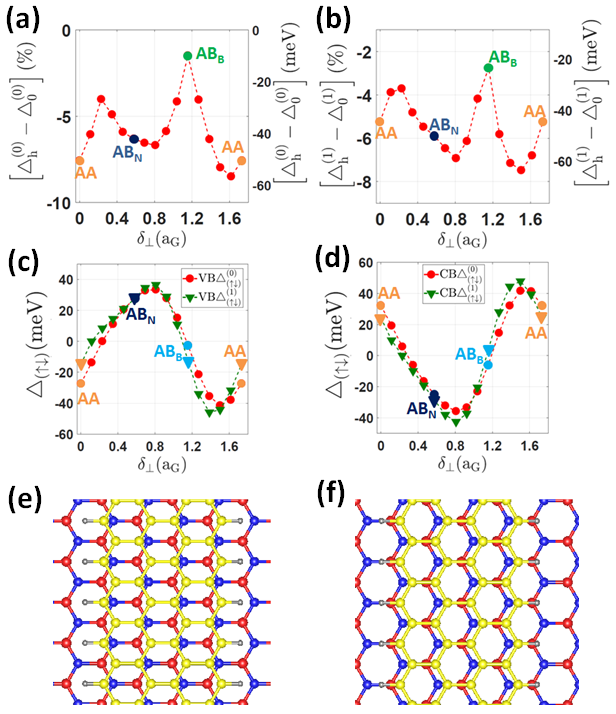}
  \caption{            
           Evolution of the band gaps and spin splittings of a ZGNR with $N=4$ placed on hBN as a function of $\delta_\perp$.
           (a), (b), Change of $\Deltaz$, $\Deltao$, respectively, due to the presence of the hBN.
           (c), (d) Spin splitting $\Deltasp$, at $k=\pi/\azgnr$, and close to $\Deltaz$, due to the presence of hBN for the VB and CB, respectively.
           (e), (f) Stacking configuration corresponding to the values of $\delta_{\perp}$ for which the spin splitting
           $\Deltasp$ is maximized, shown in ~(c),~(d): $\delta_\perp = 0.8 a_G$ in (e), and $\delta_\perp = 1.5 a_G$ in (f).
         } 
  \label{fig:ZGNR-hBN-d_perp}
 \end{center}
\end{figure}

Figure~\ref{fig:ZGNR-hBN-d_perp} shows the effects on the ZGNR band structure of a shift along
the ribbon's transverse direction away from the $AA$ stacking configuration.
We see that the reduction of $\Deltaz$ and $\Deltao$ oscillates with $\delta_\perp$,
Fig.~\ref{fig:ZGNR-hBN-d_perp}~(a),~(b).
The spin splitting also oscillates with $\delta_\perp$
Fig.~\ref{fig:ZGNR-hBN-d_perp}~(c),~(d), in  a very similar way both around $\Deltaz$ and $\Deltao$ for valence and conduction band. As for the band-gap the effect of the hBN on the spin splitting is minimal
for the $AB_B$ stacking configuration. Also for values of $\delta_\perp$ such that a configuration between
AA and $AB_N$ is realized the spin splitting can be tuned very close to zero.
We find that by varying $\delta_\perp$ the Zeeman splitting is maximized when a configuration close to the $AB_N$  
stacking $\left(\delta_\perp = 0.8 a_G \right)$ or not too far from the AA stacking one $\left(\delta_\perp = 1.5 a_G \right)$. For these configurations the spin splitting is about 40~meV.
Fig.~\ref{fig:ZGNR-hBN-d_perp}~(e),~(f) show the stacking configurations corresponding to 
$\delta_\perp = 0.8 a_G$ and $\delta_\perp = 1.5 a_G$, respectively. We see that in both cases the carbon atoms of one of the GNR sublattices
are very close to the nitrogen atoms whereas the carbon atoms of the other sublattice are very close to the
boron atoms. Due to the details of the electrostatic environment created by the hBN we conclude that these, among the configurations that
we have considered, are the ones that maximize the breaking of the ZGNR sublattice symmetry and therefore the spin splitting
of the spin polarized edge modes.

\begin{figure}[htb]
 \begin{center}
  \includegraphics[width=\columnwidth]{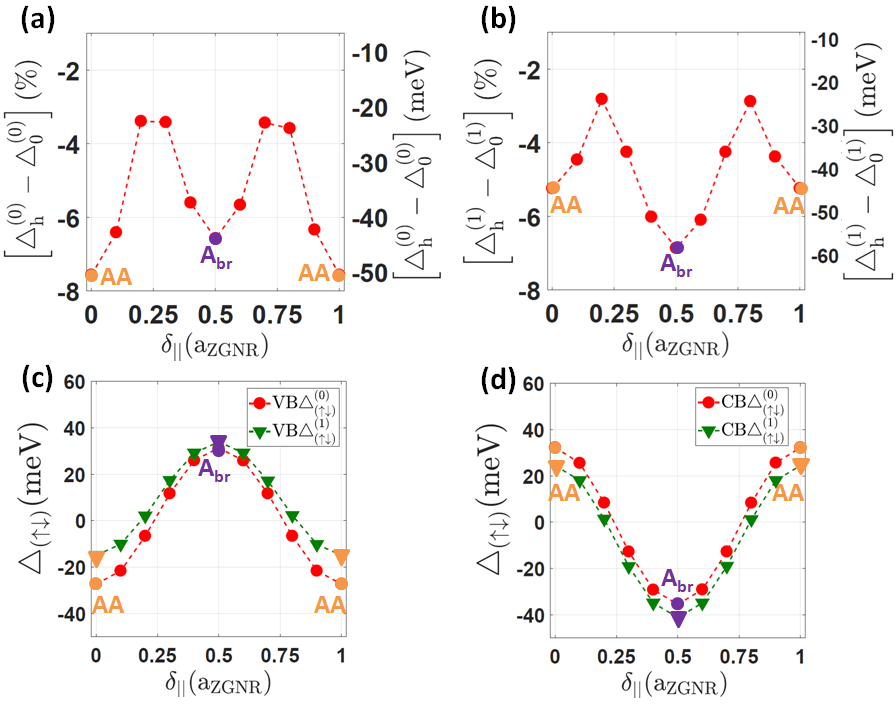}
  \caption{
           Evolution of the band gaps and spin splittings of a ZGNR with $N=4$ placed on hBN as a function of a 
           shift $\delta_\parallel$ away from AA stacking.
           (a) Change of $\Deltaz$ due to the presence of the hBN.
           (b) Change of $\Deltao$ due to the presence of the hBN.
           (c), (d) Spin splitting $\Deltasp$, at $k=\pi/\azgnr$, and close to $\Deltaz$, due to the presence of hBN for the CB and VB respectively.
         } 
  \label{fig:ZGNR-hBN-d_parallel}
 \end{center}
\end{figure}

Figure~\ref{fig:ZGNR-hBN-d_parallel} shows how the band gap and the spin splitting change by shifting the ZGNR
away from the AA stacking along the longitudinal direction. As for the case of a perpendicular shift,
we see that both the gap and the spin-splitting oscillate with $\delta_\parallel$. 
Both the gaps, $\Deltaz$ and $\Deltao$, and the spin splitting are symmetric with respect to 
$(\delta_\parallel-(1/2)a_{\rm ZGNR})$
This can be understood considering that for 
$\delta_\parallel=(1/2) a_{\rm ZGNR}$ we obtain the \abr configuration and that shifts along the longitudinal direction 
around such configuration lead to equivalent stackings.
The results of \ref{fig:ZGNR-hBN-d_parallel}~(b)-(d) show that for the  \abr configuration, see Fig.~\ref{fig:gnrs}~(f), both 
$\Deltao$ and the spin splitting are maximized.
Our results show that, due to the details of the electrostatic
potential created by the atoms forming the heterostructure, the strongest sublattice-breaking effect of hBN is not obtained
for the $AA$ stacking configuration, as one would naively expect, but for configurations as the ones shown in Fig.~\ref{fig:gnrs}~(f) and 
Fig.~\ref{fig:ZGNR-hBN-d_perp}~(e),~(f) in which the carbon atoms are slightly off from being directly
above the nitrogen and carbon atoms.

\begin{figure}[htb]
 \begin{center}
  \includegraphics[width=\columnwidth]{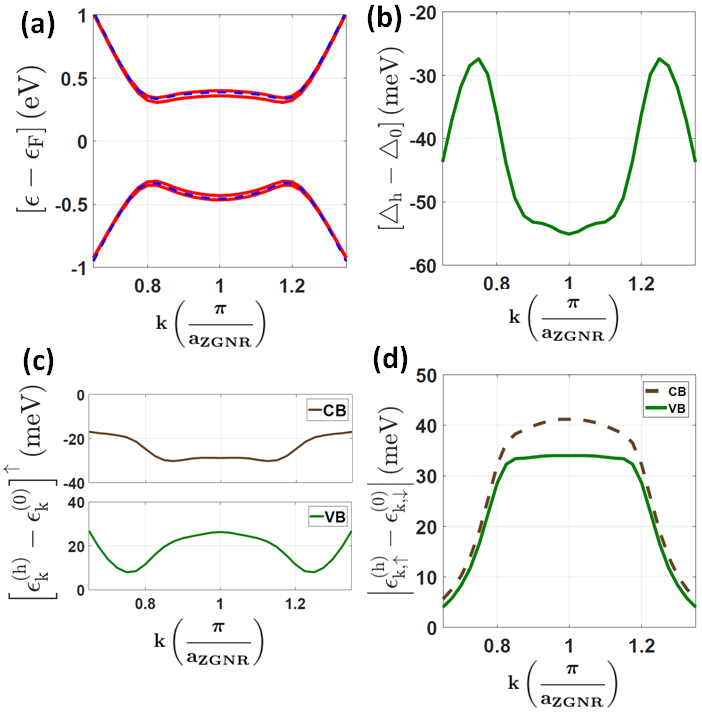}
  \caption{     
           Results for a ZGNR with $(N=4)$ placed on hBN in the  $A_{br}$ stacking configuration. 
           (a) Band structure, 
               the dashed lines show the bands for the isolated ZGNR.    
           (b) Difference, for $k$ close to $\pi/a_\mathrm{ZGNR}$, between the band gap of the hBN-ZGNR heterostructure, $\Delta_h$,
               and the band gap of the isolated ZGNR $\Delta_0$.
           (c) The top panel shows the difference for $k$ close to $\pi/a_\mathrm{ZGNR}$ between the ZGNR-hBN heterostructure's CB
               and the isolated ribbon's CB for the AA stacking configuration. 
               The bottom panel show the difference between the VBs.
           (d) Spin splitting as a function of $k$ for the ZGNR-hBN heterostructure's CB and VB.
         }
  \label{fig:ZGNR-hBN-Abr}
 \end{center}
\end{figure}

Figure~\ref{fig:ZGNR-hBN-Abr} shows the low-energy band structure of ZGNR-hBN for the \abr configuration.
As to be expected we see, Figure~\ref{fig:ZGNR-hBN-Abr}~(c), that the spin splitting induced by the presence of hBN
decreases as we move away from the $k=\pi/a_{\rm ZGNR}$ point, i.e as we move away from the value of $k$ for which the 
locking of the spin and sublattice degree of freedoms for the edge states is the strongest.

\begin{figure}[htb]
 \begin{center}
  \includegraphics[width=\columnwidth]{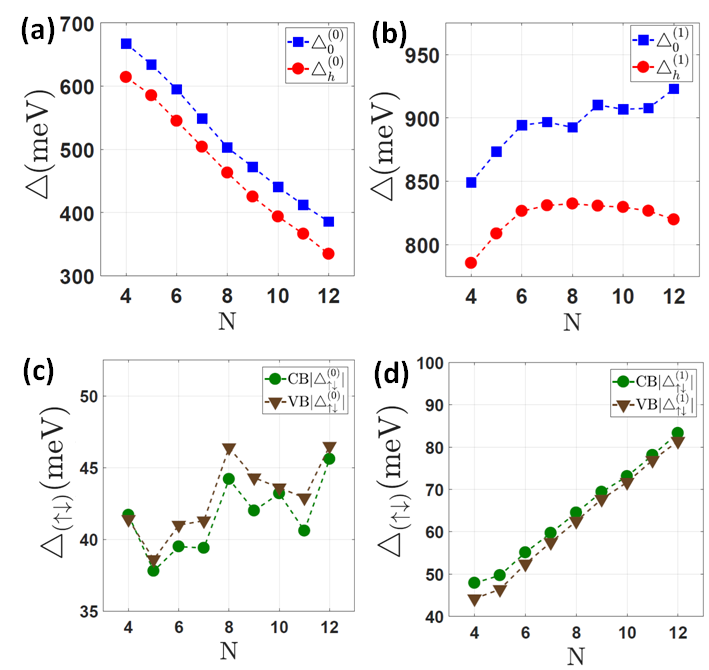}
  \caption{       
           Effect of the ribbon width, $N$ for a ZGNR-hBN heterostructure with
           stacking configuration shown in Fig.~\ref{fig:ZGNR-hBN-d_perp}~(f) corresponding to $\delta_{\perp}=1.5 a_G$, 
           value of $\delta_\perp$ for which the spin splitting $\Deltasp$ is maximized.
           $\Deltaz$, (a), and $\Deltao$, (b), as a function of $N$ for the ZGNR-hBN heterostructure and the isolated ribbon.
           $\Deltasp$ for CB and VB around the $X$ point, (c), and the $k=\pi/\azgnr$, (d).
         }
  \label{fig:ZGNR-hBN-N}
 \end{center}
\end{figure}

The results of Figs.~\ref{fig:ZGNR-hBN-d_perp},~\ref{fig:ZGNR-hBN-d_parallel} show that by shifting the ZGNR away from the $AA$ configuration
we have the maximum spin splitting for shift in the transverse direction with $\delta_{\perp}=1.5 a_G$.
It is then interesting to see how the main features of the band structure of a ZGNR-hBN system with $\delta_{\perp}=1.5 a_G$
vary as we change the width of the nanoribbon. The results are shown in Fig.~\ref{fig:ZGNR-hBN-N}.
For an isolated ZGNR we have that as $N$ increases the band gap $\Deltaz$ induced by the antiferromagnetic ordering of the edge states decreases, whereas $\Deltao$ remains approximately constant~\cite{son2006}. This is shown by the squares symbols in Fig.~\ref{fig:ZGNR-hBN-N}~(a), and (b), respectively.
The circles in the same figures show the results for the ZGNR-hBN heterostructure. We see that the presence of hBN does not affect qualitatively the scaling of of $\Deltaz$ and $\Deltao$ with respect to $N$.

\begin{figure}[htb]
 \begin{center}
   \includegraphics[width=\columnwidth]{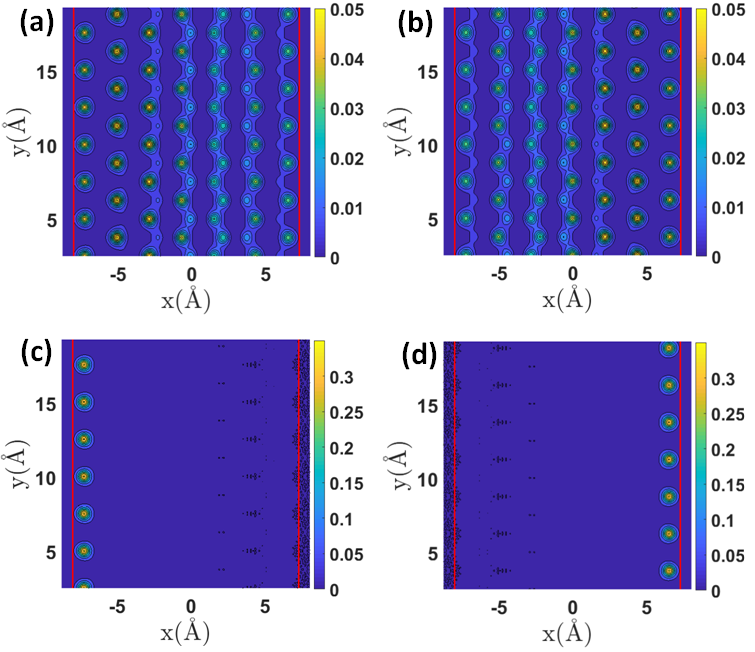}
   \caption{
     Electron charge density for a pristine ZGNR with $N = 7$. 
     (a),~(b) ((c),~(d)) show the electron density of spin up and spin down states, respectively close to $\Deltaz$ ($k = \pi/\azgnr$). 
            }
   \label{fig:charge_density}
  \end{center}
 \end{figure} 

It is then interesting to see how the spin splitting induced by the presence of hBN scales with $N$.
Fig.~\ref{fig:ZGNR-hBN-N}~(c),~(d) show the spin splitting around $\Deltaz$ and $\Deltao$, respectively.
Contrary to $\Deltaz$ the spin splitting around it depends very weakly on $N$.
This can be qualitatively understood considering that the states close to $\Deltaz$ 
are not strongly localized at the edges as shown in Fig.~\ref{fig:charge_density}~(c),~(d)
and their localization does not change much by varying the width of the ribbon.
As a consequence, the fact that the carbon atoms at the opposite edges of the ribbons
see a different electrostatic potential being either on top of nitrogen atoms or boron atoms, 
does not cause a spin splitting that depends strongly on the ZGNR's width, as shown in Fig.~\ref{fig:ZGNR-hBN-N}~(c).
The opposite is true for the states close to $k=\pi/\azgnr$: in this case the states are strongly localized
to the edges and this localization increases with the ribbon's width enhancing the spin splitting
due to the sublattice breaking effect of hBN on the ribbon, Fig.~\ref{fig:ZGNR-hBN-N}~(d).
We therefore conclude that the semimetal character of ZGNRs placed on hBN can be increased
by considering wider ribbons.

\section{Conclusions}
\label{sec:conclusions}
%
We have studied how the presence of hBN affects the electronic structure of armchair and zigzag graphene nanoribbons.
We have obtained how hBN modifies the low energy properties of the graphene ribbons' bands and how these changes
depend on the stacking configuration.
Pristine armchair graphene nanoribbons have always a finite band gap. We find that for the class of armchair graphene
nanoribbons with the smallest band gap, ribbons of width $N=3n-1$ (with $n$ a positive integer), the presence of hBN 
can modify the GNR's gap by as much as 20\%. 
For the armchair graphene nanoribbons for which the band-width is larger when isolated, ribbons
of width $N=3n$  and $N=3n+1$, the presence of hBN modifies the size of the gap only up to about 6\%.
The effect of hBN is much more significant for zigzag graphene nanoribbons.
For these ribbons the band gap is due to the antiferromagnetic ordering of the edge states
and the fact that the carbon atoms at the opposite edges of the ribbon belong to different sublattices
implies that the presence of hBN, by breaking the sublattice symmetry, can strongly
modify the low-energy features of the ribbon.
The presence of hBN can induce a significant spin splitting of the conduction and valence band
and drive the ribbon into a half-metallic state. 
We find that such spin splitting is maximized
for the so called {\em bridge} stacking configuration in which the carbon-carbon links of the GNR 
cross the boron-nitride links of hBN and for configurations close to the $AA$ stacking configuration,
but not for the $AA$ stacking configuration itself.
For a zigzag GNR of width $N=4$ we find that the spin splitting of the conduction and valence
bands can be maximized, by varying the stacking configuration, to about 40~meV conservatively assuming a GNR-hBN distance equal t0 3.5$\AA$.

Our results show that hBN in general modifies the low energy features of GNRs and that this effect
can be tuned to some extent by varying the stacking configuration. 
For zigzag GNRs, due the spin-sublattice locking of the edges states, the presence of hBN induces a spin
splitting of the conduction and valence bands that can be exploited, by properly doping the GNRs, to drive
the ribbon into a half-metallic state. The ability to achieve a relatively large spin splitting of the
conduction and valence bands without introducing external magnetic fields or proximity to ferromagnetic materials
could be very helpful in spintronics applications and in particular to realize quasi 1D ideal spin-filters.
In addition, by proximitizing the ribbon to a superconducting system with spin-orbit coupling, such as the surface
of Pb, it should be possible to drive a ZGNR-hBN heterostructure in a topological superconducting state
supporting Majorana modes. The possibility to realize Majoranas in GNRs is interesting given that
GNR are ideal 1D systems and therefore can easily be driven into a situation when only one spinful band is occupied
vs the case of semiconductor-superconductor nanowires where in typical experimental conditions several bands
are occupied~\cite{antipov2018}. Moreover, advances in GNRs growth make possible the realization of high quality
ribbons with designed width and therefore could be promising systems to realize Majoranas nanowires networks,
a necessary step to realize a Majorana based topological quantum bits.

\section{Acknowledgments}

We thank Eric Walters for discussions. 
This work was supported by 
NSF Grant No. DMR-1455233, and ONR-N00014-16-1-3158.
The numerical calculations have been performed on computing facilities at William \& Mary which were provided by contributions from the National Science Foundation, the Commonwealth of Virginia Equipment Trust Fund, and ONR.




%


\end{document}

%% file: mycommands.tex
\usepackage{graphics, bm, xspace}
\usepackage{graphicx}
\usepackage{psfrag}
\usepackage{amsmath}
\usepackage{amssymb}
\usepackage{epsfig}
\usepackage{subfigure}
\usepackage{grffile}
\usepackage{times}
\usepackage{color}
\usepackage{multirow}
\usepackage[bookmarks=false,linkcolor=blue,urlcolor=blue,colorlinks,citecolor=blue]{hyperref}
\usepackage{float}

\newcommand{\beq}    {\begin{equation}}
\newcommand{\enq}    {\end{equation}}

\newcommand{\df}     {\equiv}

\newcommand{\abr}     {${\rm A_{br}}$\xspace}

\newcommand{\aagnr}     {a_{\rm AGNR}}
\newcommand{\azgnr}     {a_{\rm ZGNR}}

\newcommand{\Deltaz} {\Delta^{(0)}} 
\newcommand{\Deltao} {\Delta^{(1)}} 
\newcommand{\Deltasp}{\Delta_{(\uparrow\downarrow)}}

